\documentclass[11pt,a4paper]{article}
\usepackage{jcappub}

\usepackage[utf8]{inputenc}
\usepackage{graphicx}
\usepackage{amssymb}
\usepackage{amsmath}
\usepackage{units}
\usepackage{listings}
\usepackage{rotating}
\usepackage{multirow}
\usepackage{aas_macros}
\usepackage{color}
\usepackage{ulem}
\graphicspath{%
  {./figs/},%
}

\newcommand{\mysim}{\sim\!}

\newcommand{\bpm}{\begin{pmatrix}}
\newcommand{\epm}{\end{pmatrix}}
\newcommand{\OM}{\Omega_{\textrm{M}}}

\newcommand{\OL}{\Omega_{\Lambda}}
\newcommand{\sint}{\sigma_{\textrm{int}}}
\newcommand{\cred}{\chi_{\textrm{red}}}

\title{The Self-Calibrating Hubble Diagram}
\author{Ulrich Feindt,}
\author{Marek Kowalski}
\author{and Kerstin Paech}

\affiliation{Physikalisches Institut, Universit\"{a}t Bonn, Nu\ss allee 12,
53115 Bonn, Germany}

\emailAdd{feindt@physik.uni-bonn.de}
\emailAdd{kowalski@physik.uni-bonn.de}
\emailAdd{paech@physik.uni-bonn.de}

\abstract{ As an increasing number of well measured type Ia supernovae
  (SNe~Ia) become available, the statistical uncertainty on $w$ has
  been reduced to the same size as the systematic uncertainty. The
  statistical error will decrease further in the near future, and
  hence the improvement of systematic uncertainties needs to be
  addressed, if further progress is to be made. We study how
  uncertainties in the primary reference spectrum -- which are a main
  contribution to the systematic uncertainty budget -- affect the
  measurement of the Dark Energy equation of state parameter $w$ from
  SNe~Ia. The increasing number of SN observations can be used to
  reduce the uncertainties by including perturbations of the reference
  spectrum as nuisance parameters in a cosmology fit, thus
  ``self-calibrating'' the Hubble diagram.\\
  We employ this method to real SNe data for the first time and find
  the perturbations of the reference spectrum consistent with zero at
  the 1\%-level. For future surveys we estimate that $\mysim 3500$ SNe
  will be required for our method to outperform the standard method of
  deriving the cosmological parameters.
}

\begin{document}

\maketitle
\flushbottom

\section{Introduction}
\label{sec:introduction}

Since the discovery of the accelerated expansion of the universe
\cite{perlmutter1998,garnavich1998,schmidt1998,riess1998,perlmutter1999}
type Ia supernovae (SNe Ia) have proven an excellent tool for probing
the Dark Energy equation of state (EOS). Over the last decade the
sensitivity of the measurements has increased due to the inclusion of
larger, better calibrated data sets
\cite{knop2003,astier2006,woodvasey2007,kowalski2008,hicken2009b,kessler2009,%
  amanullah2010,guy2010}. By now, the statistical uncertainties in the
measurement of the EOS match the size of the systematic ones (see
e.g.~\cite{amanullah2010,conley2011}). With an expected increase in
the number of SNe that will be available from future surveys, the
statistical errors will decrease fast, hitting the ``systematic error
wall'' -- if the systematic uncertainties cannot be reduced.

Measuring cosmological parameters from SNe Ia relies on comparing the
luminosity distances of distant SNe with those of nearby ones. While
the measurements of high-redshift SNe rely on redder bands (e.g.\ $r$,
$i$ and $z$), small/intermediate distances require bluer bands (e.g.\
$u$ and $g$) as well~\cite{regnault2009}. Therefore the chromatic
error in the flux calibration is a main contribution to the systematic
uncertainties of cosmological parameters. Two main sources for
chromatic uncertainties are the systematic errors for the zero-point
determination for any given survey and the measurement of the
fundamental flux standard that all surveys use to compare their
measurements.

For the effect of zero-point uncertainties, it was
shown~\cite{amanullah2010,conley2011} that a 1\% shift in the
zero-point for a given high-redshift survey causes a variation of
$\mysim 0.04$ in the dark energy equation of state parameter $w$.

With regard to the error for the fundamental flux standard, the
generic calibration procedure can be summarized as follows: One or
more fundamental spectroscopic standard stars, for which the spectral
energy distributions (SEDs) are assumed to be known from careful
modelling, serve as a reference. For example, Bohlin~\cite{bohlin2004}
observed three white dwarfs (WDs) for this purpose using HST,
confirming repeatability as well as consistency (both at sub-0.5\%
level) with model SEDs. Subsequently, a set of fundamental standard
stars - among them Vega and BD+17$^{\circ}\!$4708 - was observed with
the same instrument, adopting the calibration from the WDs. In a final
step (undertaken by the ``calibration'' user), a system of secondary
standard stars, e.g.\ that of Landolt \cite{landolt1992}, for which
the relative calibration to the fundamental standard star is precisely
determined, is used to determine the nightly calibration of the survey
telescopes. Each step in the calibration chain will introduce a new
set of errors. For photometric surveys, the detailed pass bands need
to be known in addition and introduce further uncertainties. As the
calibration error is a main component of systematic uncertainties,
cosmology oriented surveys have undertaken great efforts to achieve
calibration uncertainties below 1\%, notably the Sloan Digital Sky
Survey (SDSS)~\cite{ivezic2007,padmanabhan2008} and
SNLS~\cite{regnault2009}. Less attention is given to the uncertainty
due to the primary reference star, which is believed to be below 1\%
as well, but is ultimately model dependent.\\

For uncertainties of the fundamental flux standard the calibration
uncertainties can be reduced by a self-calibrating fit. In a
self-calibration approach, studied with simulated data
in~\cite{kim2006,faccioli2011}, the growing number of observed SNe is
used to constrain the uncertainties by including them as nuisance
parameters in the fit. This is possible because SNe Ia are observed in
multiple bands and are standardizable, i.e.\ the absolute magnitude is
very similar for all SNe after standardization. For instance, one can
relate the $r$ band response for $z\approx0.5$ to the $g$ band
response of a SN at $z\approx0.2$. Because for each SN multiple bands
are available, and their response has a distinct redshift dependence,
one can break a potential degeneracy with achromatic cosmological
effects.
 
The main focus of this paper is to constrain the uncertainty of the
fundamental flux standard using a self-calibrating fit. In doing so,
the error wall can be avoided as the statistical uncertainties will
continue to decrease as more SN Ia data become available. A
self-calibrating fit can in principle constrain the uncertainties in
both hardware and reference star calibration given large enough data
sets. In~\cite{kim2006,faccioli2011} zero-point uncertainties were
added to the apparent magnitudes in each observed filter and it was
shown that while the advantages of self-calibration are not evident
for the currently available amount of data, they will become more
significant for future surveys.
  
In this work we apply the method of self-calibration to real SN Ia
data for the first time to search for systematic errors in the
reference spectrum BD+17$^{\circ}\!$4708. For our analysis we use
light-curve data from the Union2 compilation \cite{amanullah2010},
extend by the SNLS 3~year data \cite{guy2010}, which we fit using the
SALT2 light-curve fitter~\cite{guy2007}. We then perturb the reference
spectrum of BD+17$^{\circ}\!$4708 in several wavelength bands and redo
the light-curve fitting process for each perturbation. From that we
determine the behaviour of each SN under small changes in the flux of
the reference spectrum in a given wavelength range. We then include
these dependencies on the perturbations in a cosmological fit by
introducing the size of the perturbations as a nuisance parameter for
each band. Using the data at hand, our results show that we can
constrain the calibration uncertainty to about 1\% in the $r$- and
$i$-band, and about 2\% for the $g$- and $z$-band (we do not perturb
the $u$-band due to the scarcity of available data). Finally, we
ascertain that the self-calibrating fit behaves in a statistically
sound fashion and give an estimate of the number of SNe where
self-calibration starts to outperform the standard method.

The paper is organized as follows. In section~\ref{sec:method} we
discuss our method by comparing the standard SN cosmology fit with our
self-calibrating fit. The data we used is reviewed in
section~\ref{sec:data}. The results and their implications
for future surveys are discussed in section~\ref{sec:results}.

\section{Method}
\label{sec:method}
Our analysis starts from the standard method to derive the
cosmological parameters from SN Ia measurements, which we will
summarize briefly. Subsequently, we will describe our approach to
self-calibrate the Hubble diagram and account for uncertainties in the
fundamental flux standard BD+17$^{\circ}\!$4708.

Currently two empirical relations are used to standardize SNe Ia: a)
the width-luminosity (or brighter-slower) relation, i.e.\ brighter SNe
having a slower decline
rate~\cite{phillips1993,riess1995,perlmutter1997} and b) the
brighter-bluer relation which is mainly attributed to dust absorption
but may in part also account for an intrinsic
variation~\cite{tripp1998,riess1996a,guy2005,guy2007,guy2010,
  snf2011}.

In our analysis we use the latest version of the
SALT2~\cite{guy2007} light-curve fitter which was used by
SNLS3~\cite{guy2010}. It describes the measured light-curves using
three parameters: the apparent magnitude at peak brightness $m_{\rm
  B}$, the light-curve shape expressed by generic parameter $x_{1}$,
and the colour parametrized by $c=(B-V)_{\textrm{MAX}}-\langle
B-V\rangle$. The SALT2 parameters are then used to estimate the
distances (e.g.~\cite{astier2006,guy2007,amanullah2010}) by applying
the following linear correction:
\begin{align}
  \label{eq:2}
  \mu&=m_{B}-\mathcal{M}+\alpha\cdot x_{1}-\beta\cdot c \nonumber\\
  &=\boldsymbol{V}^{T}\boldsymbol{X}-\mathcal{M}
\end{align}
with
\begin{equation*}
  \boldsymbol{V}=\bpm-\beta\\1\\\alpha\epm,\quad
  \boldsymbol{X}=\bpm c\\m_{B}\\x_{1}\epm,
\end{equation*}
where $\alpha$, $\beta$ and the magnitude $\mathcal{M} = M + 5
\log(H_0/70 \textrm{ km s}^{-1}\textrm{ Mpc}^{-1})$ are nuisance
parameters that are determined by minimizing the Hubble residuals
while fitting for the cosmological parameters. Since $H_0$ and $M$ are
degenerate we fix the Hubble constant to $H_0 = 70$~km~s$^{-1}$
Mpc$^{-1}$ for our analysis.

These corrections do not standardize the SNe perfectly and some
"intrinsic scatter" $\sint$ remains originating from gaps in our
understanding of SN physics and systematic measurement
uncertainties and has to be included in the uncertainties
on $\mu$. In the standard approach to determine the cosmological
parameters $\OM$, $\OL$, $w$, the measured distance moduli are
compared to the predictions from the cosmological model by using a
standard $\chi^2$ fit with
\begin{equation}
  \label{eq:5}
  \chi^{2}=\sum\limits_{\textnormal{SNe}}\frac{|\boldsymbol{V}^{T}
    \boldsymbol{X}-\mathcal{M}-\mu_{\textrm{cosmo}}(z;\OM,\OL,w)|^{2}}{
    \boldsymbol{V}^{T}\boldsymbol{C}(\boldsymbol{X})\boldsymbol{V}
    +\sint^{2}}  
\end{equation}
where $\boldsymbol{C}(\boldsymbol{X})$ is the covariance matrix of
$\boldsymbol{X}$ and $\sint$ is adjusted so
$\cred^{2}=\chi^{2}/({\rm d.o.f.}) = 1$. 

Measurement uncertainties of the fundamental flux standard will affect
the rest frame magnitudes depending on the redshift. At a redshift of
$\mysim 0.45$ the $r$ band corresponds to rest frame $B$,
whereas at redshift $\mysim 0.75$ this is the case for $i$.
Hence the fit actually compares different filters with each other
which means that uncertainties in the spectrum of the reference star
BD+17$^{\circ}\!$4708 will cause a large systematic uncertainty in the
resulting cosmological parameters.

One way to address this problem is to use the data itself to correct
for these uncertainties by introducing perturbations of the reference
spectrum as nuisance parameters in the correction of the distance
moduli (eq.~(\ref{eq:2})) and the subsequent fit (eq.~(\ref{eq:5})).
This way, provided with enough statistics, we can correct for
uncertainties in the reference spectrum.

\begin{figure}[t]
  \centering
  \includegraphics[width=\textwidth]{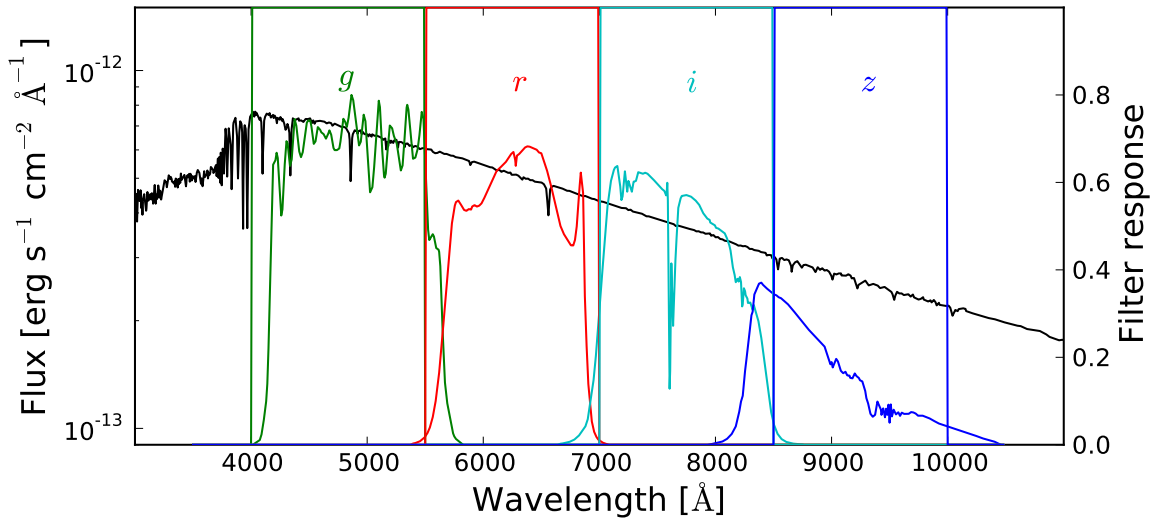}
  \caption{Perturbation bands compared to the MegaCam response
    functions~\cite{regnault2009} and the spectrum of
    BD+17$^{\circ}\!$4708 (CALSPEC, as used by SALT2~\cite{guy2007})}
  \label{fig:perturb}
\end{figure}

We start by determining how much the SALT2 parameters will be affected
by a given modification in the reference spectrum
BD+17$^{\circ}\!$4708. This is done by refitting the SN Ia light curve
data using a spectrum of BD+17$^{\circ}\!$4708 which was perturbed
using a step function, i.e.\ by increasing the flux by a factor
$(1+\delta_{j})$ in four different wavelength ranges (one at a time)
which roughly correspond to the wavelength coverage of the MegaCam
$ugriz$ filter set shown figure \ref{fig:perturb}. As an alternative
to using step functions for the perturbation of the SED of
BD+17$^{\circ}\!$4708, we also tested perturbation bands with smooth
edges. We replaced the sharp edge of the step functions with parabolic
flanks of 250~\AA\ width. The results for these perturbation bands
differ only slightly from those for step functions. However, the
functions avoid the complications due to overlapping bands. We chose
the step functions for our default tests.

From this we determine the derivatives of the SALT2 parameters with
respect to the perturbations $\delta_i$ with second-order numerical
accuracy. We choose the MegaCam filters because a large fraction of
the SNe we used either uses them (SNLS3~\cite{regnault2009}) or uses
similar filters (e.g.\ SDSS~\cite{fukugita1996}). We make this
specific choice of the perturbation bands so the results can be
interpreted in a straight forward manner. Any other choice would be as
valid. Note again that we perturb the SED of BD+17$^{\circ}\!$4708 and
not the zero points of the filters. Therefore the SNe with photometric
data in other filter systems, e.g.\ BVRI, are affected by the
perturbation as well. In section~\ref{sec:data} the derivatives are
used as part of data quality cuts, results and behaviour of the
derivatives for the data set are discussed in
section~\ref{sec:pert-light-curve}.

We include the perturbations in the distance modulus (\ref{eq:2}) by
linearizing their effect on light-curve parameters:
\begin{equation}
  \label{eq:3}
  \mu'=\mu+ \sum\limits_{j}\frac{\partial
    m_{B}}{\partial\delta_{j}}\cdot\delta_{j}+
  \alpha\sum\limits_{j}\frac{\partial
    x_{1}}{\partial\delta_{j}}\cdot\delta_{j}-
  \beta\sum\limits_{j}\frac{\partial
    c}{\partial\delta_{j}}\cdot\delta_{j} 
\end{equation}
The cosmological fit (\ref{eq:5}) is modified to include the
perturbations as parameters by replacing $\boldsymbol{V}$ and
$\boldsymbol{X}$ with:
\begin{equation}
  \label{eq:7}
  \boldsymbol{V'}=\bpm\boldsymbol{V}\\\boldsymbol{V}\delta_{g}
  \\\boldsymbol{V}\delta_{r}\\\vdots \epm
  =\bpm -\beta\\1\\\alpha\\-\beta\delta_{g}\\\delta_{g}\\
  \alpha\delta_{g}\\\vdots\epm,
  \quad
  \boldsymbol{X'}=\bpm\boldsymbol{X}\\\partial_{g}\boldsymbol{X}\\ 
  \partial_{r}\boldsymbol{X}\\\vdots\epm 
  =\bpm c\\m_{B}\\x_{1}\\\frac{\partial c}{\partial\delta_{g}}\\
  \frac{\partial m_{B}}{\partial\delta_{g}}\\
  \frac{\partial x_{1}}{\partial\delta_{g}}\\\vdots\epm , 
\end{equation}
thus extending the set of nuisance parameters by the perturbations
$\delta_j$.

A similar technique was used in~\cite{conley2011} to assess the
systematic uncertainties due to calibration errors. However, there the
aim was not to constrain the uncertainties of the reference spectrum,
but to propagate the systematic uncertainties properly. In that
analysis the light-curve template of SALT2 was retrained for each
perturbed reference spectrum. We omit this step as training a
light-curve template is beyond the scope of this work. We note,
however, that this simplification is justified because the retraining
of the template increases the effects of the perturbations only by
$\mysim 5\%$ (A.\ Conley, private communication, 2011): the degeneracy
between our nuisance parameters and the light-curve model, most
notable the colour relations of the SN Ia light-curve, is expected to
be small since the nuisance parameters are measured in the observer
frame while the SN colour relations are given in the rest-frame.
Hence, by using the full available redshift range, this affect should
largely average out. Finally, we can test the validity of this
approximation by comparing our cosmological results for a
self-calibrating fit to some of the results found in \cite{conley2011}
which include the systematic errors from uncertainties in the
reference star colours (see section \ref{sec:cosmo-fits}).

\section{Data set and selection}
\label{sec:data}

For the method to work, the SN data set should be distributed over a
large redshift range. Therefore we chose the Union2 data set
\cite{amanullah2010} for our analysis which we expanded by 166 new SNe
from the SNLS 3 year release \cite{guy2010} using only those not
classified as SNe Ia$\star$. (SNe Ia$\star$ correspond to Confidence
Index 3 of \cite{howell2005}, i.e.\ the spectrum matches a SN Ia better
than any other SN type, but another SN type --- usually SNe Ic --- is
not ruled out from the spectrum alone.)

After fitting all light curves using SALT2 we followed the framework
used in \cite{amanullah2010} to ensure sufficient quality of the data.
All SNe are required to meet the Union2 criteria: (1) There is data
from at least two bands with rest-frame wavelength between 3000~\AA\
and 7000~\AA, i.e.\ the default SALT2 wavelength range. This is very
important for our analysis because the number of available bands
affects the precision with which the SALT2 colour is determined. (2)
There are at least five photometric measurements (3) There is at least
one measurement between -15 and 6 (rest frame) days relative to the
$B$-band maximum; (4) The fit values for $x_{1}$, including the fit
uncertainties, lie between $-5<x_{1}<5$; (5) The CMB-centric redshift
is greater than $z > 0.015$.

To this list of requirements we add our own. As described in section
\ref{sec:method}, we include the effects of the perturbation of
BD+17$^{\circ}\!$4708 to our fit by linearizing them according to
equation (\ref{eq:3}). For the linearization to be justified, we
require that the quadratic terms in the Taylor expansion and hence the
second derivatives are small. The quadratic term for e.g.\ the colour
perturbed in $r$ is $
  \label{eq:6}
  \frac{1}{2}\frac{\partial^{2}c}{\partial\delta_{r}^{2}}\cdot\delta_{r}^{2}.
$

While the light curve fits result in a first derivative that is
$\lesssim\! 1$ for all SNe as well as a similarly small second
derivative for most SNe, there are some cases for which the second
derivative is much larger, in a few cases even $>\! 1000$. Such large
second derivatives appear only for SNe with few photometric
measurements. For a typical uncertainty of $\delta\! =\! 1\%$, both
the linear and quadratic terms are the same order of magnitude for a
second derivative of about 200. We hence require of each SNe that the
absolute values of its second derivatives $|\partial^{2}
m_{B}/\partial \delta_{j}^{2}|$, $|\partial^{2} c/\partial
\delta_{j}^{2}|$ and $|\partial^{2} x_{1}/\partial \delta_{j}^{2}|$
are smaller than 200. So this cut is chosen so that the quadratic term
is at most the same size as the linear term.

After quality cuts there are 538 SNe from Union2 and 161 SNe from
SNLS3 in our data set, 11 SNe are solely excluded because of
non-linearity (9 for Union2 and 2 for SNLS3). Note that of the 569 SNe
in Union2 before the $3\sigma$-clipping, 22 were excluded beforehand.
These SNe all have photometric measurements in filter systems for which
the zero points for BD+17$^{\circ}\!$4708 could not be determined
reliably.

To clean the data from outliers, the robust $3\sigma$-clipping as in
\cite{kowalski2008} was adopted: the data set was first fit using
median statistics~\cite{gott2001} with $\mathcal{M}$ as the only free
parameter (using $\alpha=0.12$ and $\beta=2.51$ as well as a flat
cosmology with $\OM=0.281$ (values taken from \cite{amanullah2010})
with $\sint=0.1$). SNe outside a $3\sigma$ range were then excluded.
Note that median statistics were only used to exclude outliers, for
further fits we minimized $\chi^{2}$ according to eq.\ (\ref{eq:5}).
This procedure leaves 501 SNe in Union2 and 158 in SNLS3 and thus a
total of 659 SNe for our analysis. Note that considerably more SNe
were rejected than the 12 SNe in \cite{amanullah2010}. This is due to
the simplified approach undertaken here: The clipping was not done
separately for each sample (and not using different $\sint$), but for
the whole data set at once. We note that the aim of this work is the
first application of a method to incorporate systematic errors and not
a new determination of cosmological parameters.

\section{Results}
\label{sec:results}

\subsection{Light-curve fits for the perturbed reference spectrum}
\label{sec:pert-light-curve}

\begin{figure}[t]
  \centering
  \includegraphics[width=\textwidth]{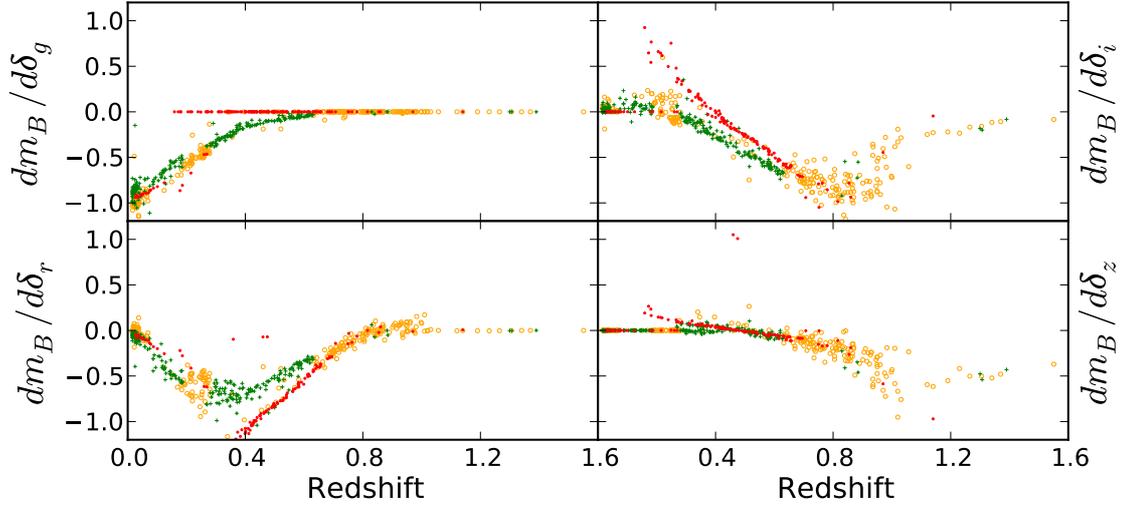}
  \caption{Derivatives of the peak magnitude $m_{\textrm{B}}$ w.r.t.
    the perturbations. The colours and marker shapes indicate the
    number of bands from which data was available and in the default
    wavelength range of SALT2 (3000 \AA\ to 7000 \AA): 2 (red dots), 3
    (orange circles), 4 or more (green crosses).}
  \label{fig:der-mB}
\end{figure}

\begin{figure}[t]
  \centering
  \includegraphics[width=\textwidth]{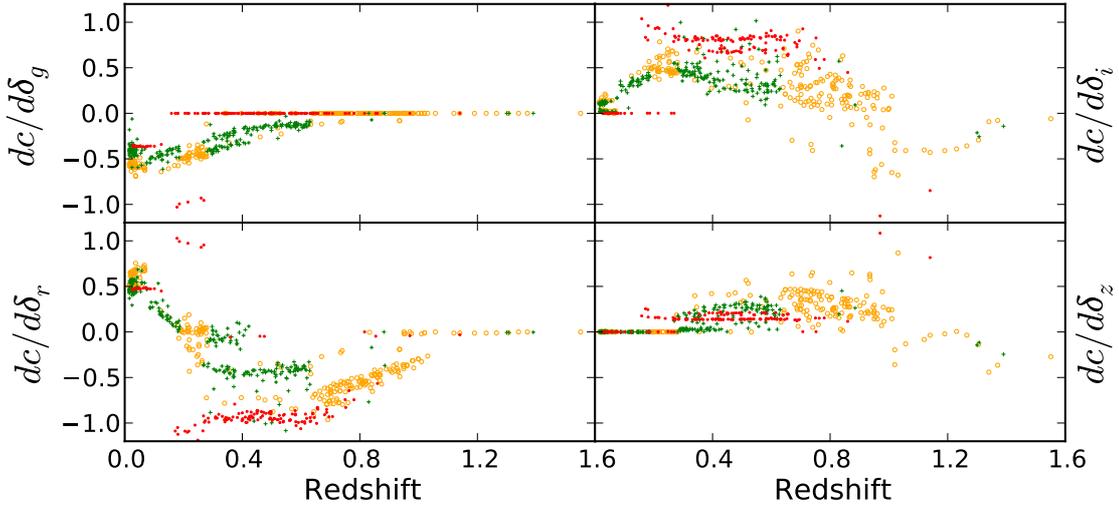}
  \caption{Derivatives of the colour $c$ w.r.t. the perturbations.
    Colours and marker as in figure \ref{fig:der-mB}.}
  \label{fig:der-c}
\end{figure}

\begin{figure}[t]
  \centering
  \includegraphics[width=\textwidth]{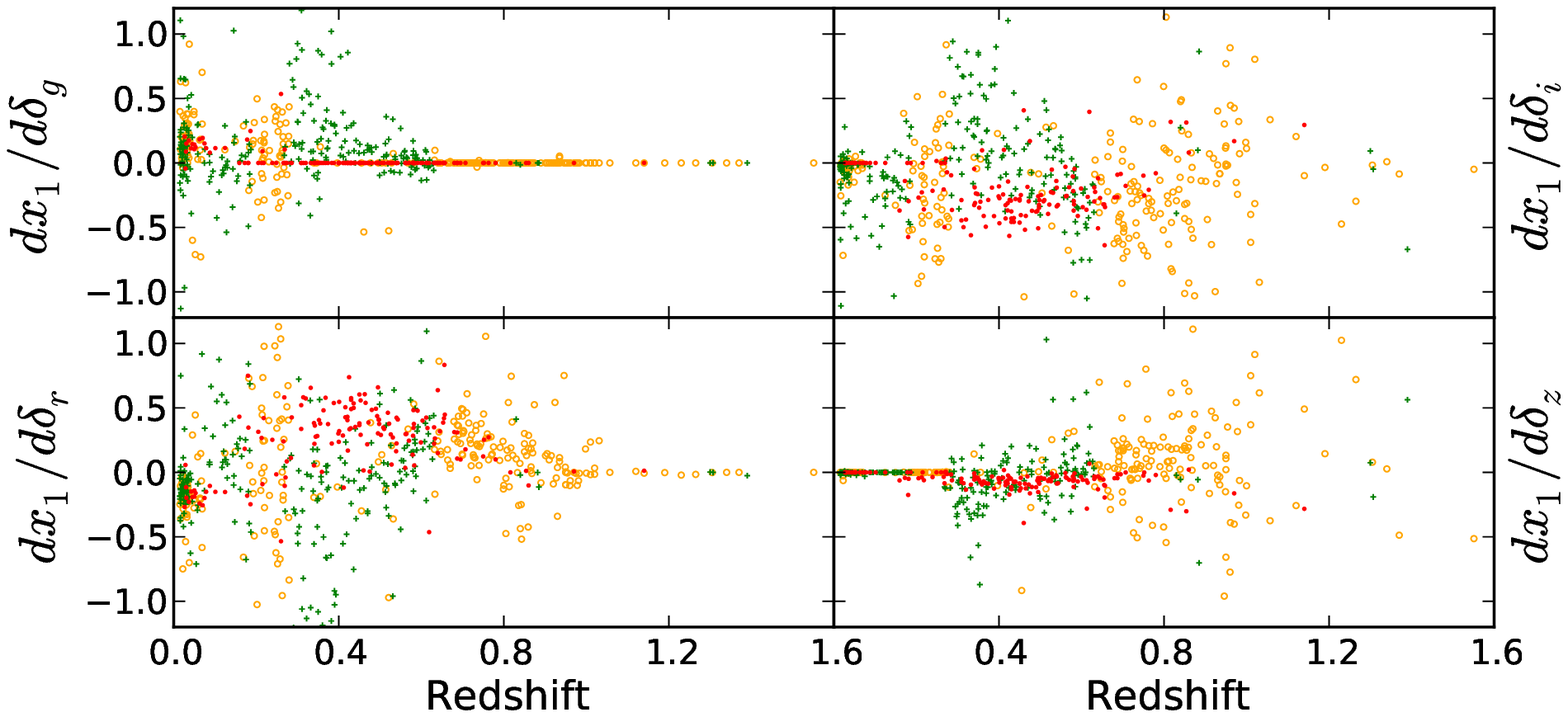}
  \caption{Derivatives of the stretch parameter $x_{1}$ w.r.t. the
    perturbations. Colours and markers as in figure \ref{fig:der-mB}.}
  \label{fig:der-x1}
\end{figure}

The behaviour of the light-curve fit results under perturbation of
BD+17$^{\circ}\!$4708 differs greatly from SN to SN. This is expected,
as the effects of the perturbation depend on the redshift of the SN as
well as the number of bands for which data is available. Figures
\ref{fig:der-mB}--\ref{fig:der-x1} show the derivatives of the
light-curve parameters with respect to the perturbations in a given
band. We also tested perturbations of 2\% and 3\%, but the change of
the derivatives is insignificant. Hence we used derivatives calculated
form 1\% perturbations throughout our analysis. A general behaviour
can be observed that matches the expectations well. Firstly, there is
a redshift dependence of the effects: Perturbations in $g$ have the
greatest effect on low-redshift SNe, whereas perturbing $r$ and $i$
affects intermediate redshifts and $z$ takes effect for high
redshifts. Secondly, the number of available bands influences the
strength of the effect. Generally the perturbation will cause a
smaller change in the fit parameters if the number of bands is larger.

Increasing the flux of BD+17$^{\circ}\!$4708 causes the \textit{peak
  magnitude} $m_{\textrm{B}}$ to drop for most of the SNe because the
light-curve points are interpreted as being brighter. This means that
the derivative $\partial m_{B}/\partial \delta_{j}$ (see figure
\ref{fig:der-mB}) is negative and its value is expected to be between
-1 and 0. We also see the aforementioned dependence on the number of
bands: when perturbing $r$, the drop in $m_{B}$ is larger if only two
bands are available because a greater fraction of light-curve points
are affected by the perturbation than if more bands were given. When
perturbing in $g$ on the other hand, most SNe with two bands are
unaffected by the perturbation since these SNe only have photometric
measurements in $R$ and $I$ which do not overlap with $g$.

The \textit{SALT2 colour parameter} $c$ shows a similar behaviour.
Hence values for $\partial c/\partial \delta_{j}$ (see figure
\ref{fig:der-c}) are expected to be between -1 and 1 with a clear
redshift dependence. We observe that the effects are very similar for
adjacent bands but have opposite signs and have the same dependence on
the number of bands as $\partial m_{B}/\partial \delta_{j}$. These
effects are best explained with a simple example: at redshift $z=0.45$
$r$ roughly correspond to the $B$ band, hence perturbing the spectrum
of BD+17$^{\circ}\!$4708 by 1\% in $r$ decreases $B$ by 0.01 which in
turn decreases $B-V$ by 0.01. Perturbing $i$ on other hand would
decrease $V$ and thus increase $B-V$. So far in this analysis we
assumed that only one colour (i.e.\ two bands) is available but if
$U-B$ is also measured, it will increase when $r$ is perturbed and
lessen the total change in $c$. Additionally we notice a sudden drop
in $\partial c/\partial\delta_{r}$ around $z\mysim 0.63$ which
coincides with a decrease of the number of bands from four to three.
These SNe are all from SNLS. At $z\gtrsim 0.63$ the effective
wavelength of the MegaCam-$g$ band corresponds to a rest-frame
wavelength below 3000 \AA\ and hence is ignored by SALT2.

An effect on the \textit{shape parameter} $x_{1}$ can also be
observed. However, this effect is more complicated than the effect on
the other parameters, as the effect of the perturbations on the shape
of the light-curve depends strongly on the number of light-curve
points that were measured and on how the measurements are distributed
relative to the peak.

\begin{figure}[t]
  \centering
  \includegraphics[width=.8\textwidth]{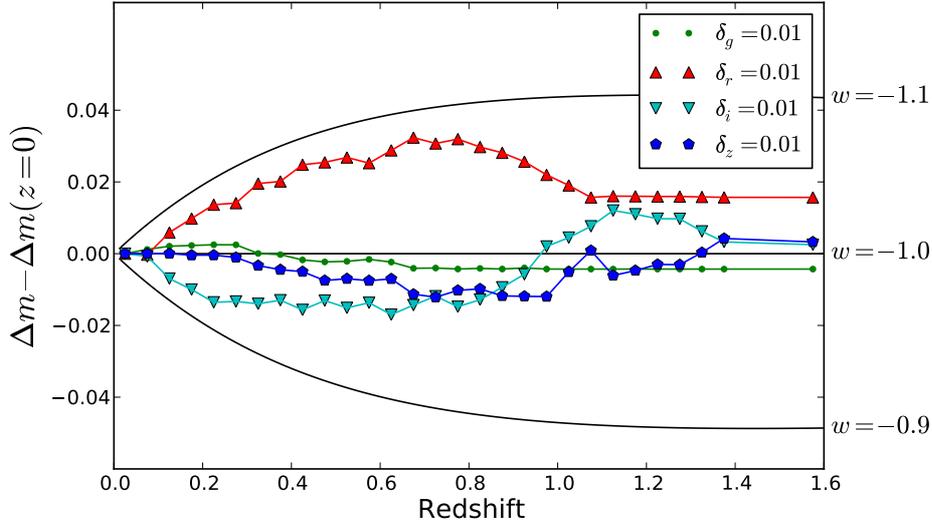}
  \caption{Change of the corrected magnitude for a 1\% perturbation in
    a given filter. Note that the values were shifted such that the
    lowest redshift bin is at zero. The values of $\Delta m(z=0)$ are
    0.004, -0.016, -0.002 and 0 for $g$, $r$, $i$, and $z$,
    respectively.}
  \label{fig:shift}
\end{figure}

We can now combine the terms according to equation (\ref{eq:3}), using
$\alpha=0.127$ and $\beta=2.802$. These values were determined in a
cosmological fit of the data as described in section
\ref{sec:cosmo-fits}. Figure \ref{fig:shift} shows the total effect of
perturbations of 1\% on the corrected magnitude. We averaged the
values in redshift bins of the size $\Delta z =0.05$. Note that the
values were shifted by $\Delta m_{g}=0.004, \Delta m_{r}=-0.016$ and
$\Delta m_{i}=-0.002$ such that they meet at $z=0$. This is justified
because cosmological parameters are anchored by nearby SNe. Shifting
the magnitudes of nearby SNe changes results for the absolute
magnitude of SNe Ia and hence changes $\mathcal{M}$ in a cosmology fit
by a similar amount. For perturbations in the $z$-band no such shift
was necessary as low-redshift SNe are not affected by the
perturbation.

The largest effect is seen in $r$ and $i$ which is expected because
all SNe are affected by perturbing those bands. For $g$ and $z$ some
SNe have no measurements that are affected by the perturbation.
Further the plot shows $\Delta m$ for different values of $w$ around a
fiducial flat wCDM model with $\OM=0.281$ and $w=-1$. This shows that
changing $w$ by 0.1 changes $\Delta m$ by 0.04 for redshifts above
$z\gtrsim 0.6$. The effect of perturbing $r$ alone is $\Delta m\mysim
0.03$. Hence we can predict a systematic uncertainty of $\mysim 0.075$
in $w$ for a 1\% uncertainty in $r$ alone.

\subsection{Cosmological fits}
\label{sec:cosmo-fits}

\begin{table}[t]
\centering
\begin{tabular}{ccccccccc}
&&$\chi^{2}$&$w$&$\mathcal{M}$& $\delta_g$&$\delta_r$&$\delta_i$&$\delta_z$ \\ \hline
\multirow{1}{*}{\begin{sideways}\end{sideways}} & &662.2 &-0.968(45) &-19.075(12) &-- &-- &-- &-- \\  \hline
\multirow{4}{*}{\begin{sideways}one band\end{sideways}} &$g$ &662.12 &-0.962(50) &-19.071(18) &0.006(23) &-- &-- &-- \\ 
&$r$ &661.68 &-1.027(96) &-19.090(24) &-- &0.008(11) &-- &-- \\ 
&$i$ &662.12 &-0.975(54) &-19.074(13) &-- &-- &-0.002(10) &-- \\ 
&$z$ &658.69 &-1.053(67) &-19.082(13) &-- &-- &-- &-0.036(19) \\  \hline
\multirow{6}{*}{\begin{sideways}two bands\end{sideways}} &$gr$ &660.81 &-1.069(108) &-19.089(24) &0.027(29) &0.016(15) &-- &-- \\ 
&$gi$ &662.05 &-0.969(58) &-19.071(19) &0.006(23) &-- &-0.002(10) &-- \\ 
&$gz$ &657.61 &-1.048(67) &-19.068(19) &0.025(24) &-- &-- &-0.043(21) \\ 
&$ri$ &661.58 &-1.039(103) &-19.097(33) &-- &0.011(15) &0.004(13) &-- \\ 
&$rz$ &658.31 &-1.105(110) &-19.095(25) &-- &0.007(11) &-- &-0.035(19) \\ 
&$iz$ &656.6 &-1.154(100) &-19.082(13) &-- &-- &-0.016(12) &-0.053(23) \\  \hline
\multirow{4}{*}{\begin{sideways}three bands\end{sideways}} &$gri$ &655.71 &-1.539(236) &-19.205(47) &0.156(58) &0.107(35) &0.057(21) &-- \\ 
&$grz$ &654.11 &-1.306(158) &-19.104(26) &0.072(35) &0.031(17) &-- &-0.057(22) \\ 
&$giz$ &654.76 &-1.171(102) &-19.064(19) &0.033(25) &-- &-0.019(12) &-0.067(25) \\ 
&$riz$ &655.61 &-1.097(109) &-19.044(40) &-- &-0.019(20) &-0.032(20) &-0.071(30) \\ 
\end{tabular}
\caption{Results of cosmology fits. The first line shows the standard
  fit, the blocks thereafter show self-calibrating fits for different
  numbers of perturbation bands. The results for the correction
  coefficients are $\alpha=0.127(6)$ and $\beta=2.802(58)$. Their
  values do not change significantly for self-calibrating fits.}
\label{tab:results}
\end{table}

The uncertainties of reference star colours have a large impact on the
measurements of $w$. SNe alone cannot constrain $w$ and $\OM$ very
well at the same time. Therefore we only fit $w$ while assuming a flat
universe with $\OM=0.281$. This emulates the inclusion of BAO and CMB
priors which constrain strongly $\OM$. The choice of $\OM$ is taken
from \cite{amanullah2010}. A proper full analysis was beyond the scope
of this paper. However, choosing a different $\OM$ between 0.26 and
0.3 only changes the mean value of $w$ and does not affect its
uncertainty. Recall that we are interested in the statistical
behaviour of the uncertainties and not the mean values.

First we fit the cosmology without perturbations, i.e.\ minimizing
$\chi^{2}$ as per equation \eqref{eq:5}. $\sint$ was tuned to 0.115
which yielded $\cred^{2}=1.011$. For further fits $\sint$ was kept at
this value to ensure comparability. The top row of table
\ref{tab:results} shows the resulting parameters for this fit. The
table only shows the parameters $w$ and $\mathcal{M}$, the nuisance
parameters $\alpha$ and $\beta$ were also determined in every fit but
they do not change significantly in a self-calibrating fit. For the
unperturbed fit we found $\alpha=0.127\pm 0.006$ and $\beta=2.802\pm
0.058$. The resulting value of $w$ is consistent with a cosmological
constant. The error of $w$, $\sigma_{w}=0.045$, only contains the
statistical uncertainty, not the systematic one.

\begin{figure}[t]
  \centering
  \includegraphics[width=\textwidth]{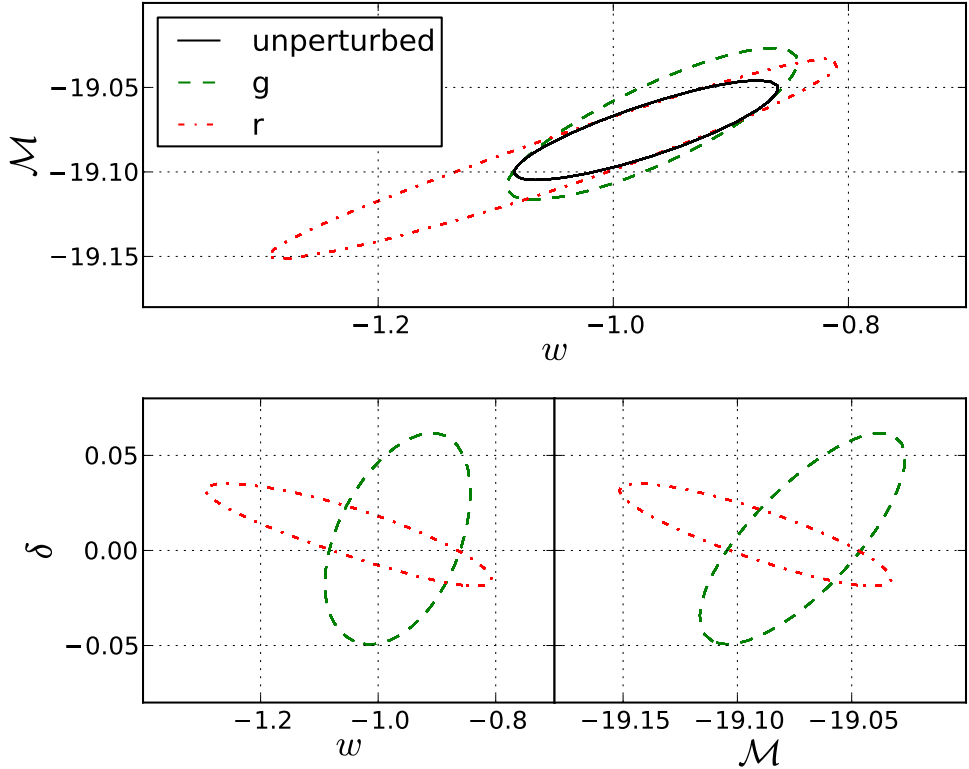}
  \caption{95\% confidence regions for self-calibration in $g$ and
    $r$ compared to the unperturbed fit}
  \label{fig:conf-gr}
\end{figure}

\begin{figure}[t]
  \centering
  \includegraphics[width=\textwidth]{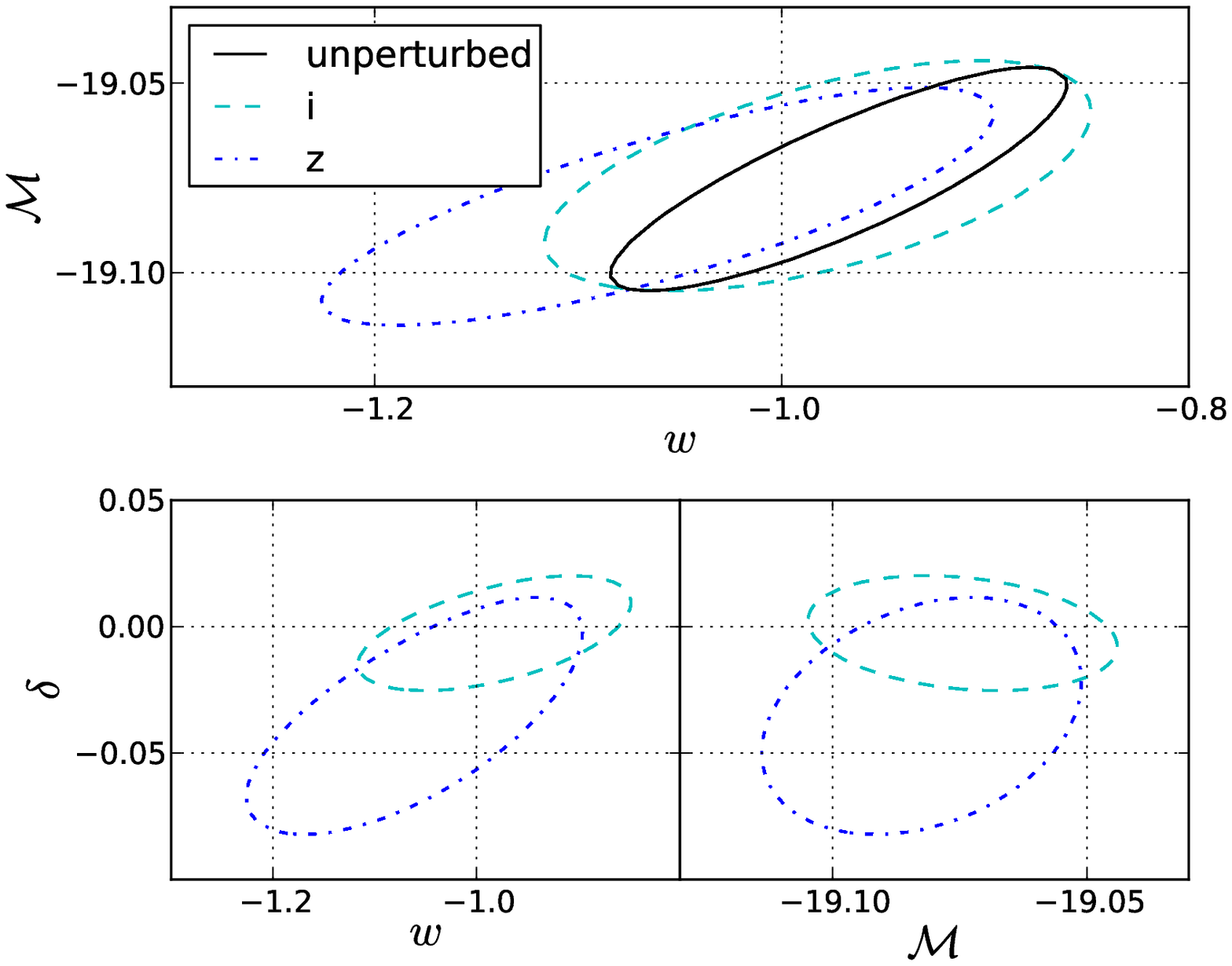}
  \caption{95\% confidence regions for self-calibration in $i$ and
    $z$ compared to the unperturbed fit}
  \label{fig:conf-iz}
\end{figure}

Next, we included a \textit{single perturbation parameter} in the fit
while leaving the other bands as they are (see the first block in
table \ref{tab:results}). The resulting deviations are of the order of
a few percent and consistent with zero. The interesting results are
the uncertainties of $w$: While perturbing the $g$- and $i$-bands
increases $\sigma_{w}$ only slightly, perturbing $r$ more than doubles
the uncertainty. Also there is a larger change in
$\sigma_{\mathcal{M}}$ for the $r$-band as well. This matches our
predictions from figure \ref{fig:shift}. The overall change of $\Delta
m$ is largest for the $r$-band and more importantly it is large for
low redshift which changes $\mathcal{M}$. The self-calibrating fit
already constrains the perturbation of the reference spectrum to about
1\% in $r$ and $i$ and to about 2\% in $g$ and $z$. It is noteworthy
that the method already constrains the flux to a level that is
compatible with the currently quoted uncertainties. The increase in
measurement uncertainty due to self-calibration is given by
\begin{equation}
  \label{eq:8}
  \Delta w =\sqrt{\sigma_{w}^{2}\textrm{(self-calibrating)}- 
    \sigma_{w}^{2} \textrm{(unperturbed)}}.
\end{equation}
This gives a $\Delta w=0.02$ for perturbations in $g$ and $\Delta w =
0.085$ for $r$. The latter matches our predictions from figure
\ref{fig:shift} well. The confidence regions for the perturbed fits
(see figures \ref{fig:conf-gr} and \ref{fig:conf-iz}) show the same
effect. While $\delta_{g}$ is less constrained than $\delta_{r}$, it
has a smaller effect on the uncertainties in $w$ and $\mathcal{M}$.

To verify that the omission of retraining the light-curve template is
justified, we can compare the $\Delta w$ to the results found by SNLS
\cite{conley2011} which include the systematic errors from
uncertainties in the reference star colour. In~\cite{conley2011} the
uncertainties of the reference colours were marginalized over to
include systematic uncertainties. Hence, this comparison is only valid
for colours which have a systematic uncertainty that is at least the
same order of magnitude as the constraints we determine using a
self-calibrating fit. The only colour that meets this requirement is
$z_{M}-V$ for which an uncertainty of 0.0178 was
used~\cite{regnault2009}. In a cosmology fit with fixed $\OM$ similar
to ours this leads to an increase of the uncertainty in $w$ from 0.058
to 0.071 which corresponds to adding 0.041 quadratically. Therefore we
find that our result for $z$ -- a quadratic increase by 0.049 -- is in
good agreement with the systematic uncertainty found
in~\cite{conley2011}. However, given that $z$ contributes only at high
redshifts and has typically a lower signal-to-noise ratio, this does
not conclusively show that the effect of retraining is insignificant.

Next, \textit{two bands} are included as perturbation parameters in
the fit, see the second block of table \ref{tab:results}. This adds
another degeneracy because one perturbation might cancel out the
others effect on the colour. For some combinations of bands ($gr$,
$gz$, $iz$) the mean values of both parameters increases. For the
other combinations ($ri$, $rz$, $gi$) they do not change, nor do the
uncertainties change much. There are two reasons for this stability.
Firstly, non-adjacent bands are more stable because increasing two
adjacent bands result in virtually no change of the colour. Secondly,
the inner bands $r$ and $i$ are better constrained and hence help to
constrain the other band.

Furthermore it is possible to include \textit{three perturbation
  parameters} (see the third block of table \ref{tab:results}), while
including all four bands would create an almost perfect degeneracy to
$\mathcal{M}$ because increasing all $\delta$s by the same amount can
the absorbed by increasing $\mathcal{M}$ as well. Hence, in our
analysis one band needs to be left unperturbed. This band should be in
the middle of the wavelength range, so only two bands are adjacent.
Thus we left the $r$-band fixed while $g$, $i$ and $z$ were perturbed.
For this fit we get $w=-1.166\pm 0.101$.

\subsection{Extrapolation to larger data sets}
\label{sec:extr-larg-data}

\begin{figure}[t]
  \centering
  \includegraphics[width=0.8\textwidth]{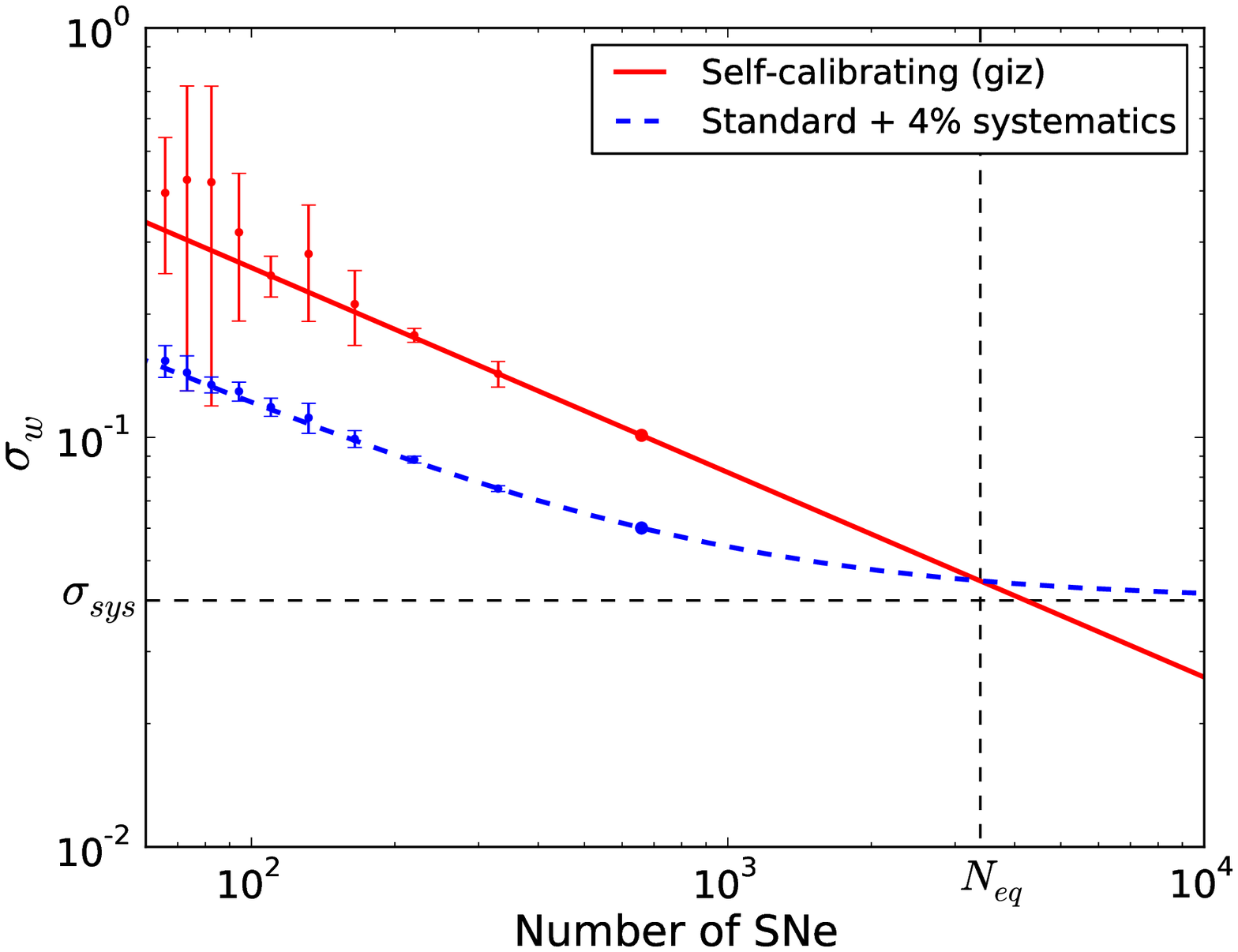}
  \caption{Extrapolated uncertainty in $w$ for the self-calibrating
    fit and the unperturbed fit with 4\% systematics. The number of
    SNe at which the self-calibrating fit is expected to be more
    precise is $N_{\textrm{eq}}\mysim 3500$. Note that the mean points
    are highly correlated as the are calculated from the same data set
    split randomly in smaller samples.}
  \label{fig:w-unc}
\end{figure}

As a next step we tested if the uncertainties behave in a
statistically sound fashion by splitting the data set randomly into
several smaller samples, i.e.\ two samples of 330 SNe, three of 220
SNe and so on. For the unperturbed fit we clearly observe that the
statistical uncertainty $\sigma_{w}$ decreases with $1/\sqrt{N}$ but
for the self-calibrating fits the uncertainties can vary greatly from
one random realization to the other. We calculated the mean
uncertainties for each sample size taking their standard deviation as
an error. Figure \ref{fig:w-unc} shows this for the unperturbed fit
and a self-calibrating fit in $g$, $i$ and $z$. Note that we add
a constant uncertainty $\sigma_{\textrm{sys}}=0.04$ to the result of
the unperturbed fit (see below). The behaviour of the uncertainties,
when compared with the $1/\sqrt{N}$ predicted scaling, confirms that
the data set behaves statistically sound in the self-calibrating fit.
   
We can extrapolate the uncertainties for larger data sets. If the
calibration does not improve, the systematic uncertainty on $w$ will
stay at the current value of about $\sigma_{\textrm{sys}}=0.04$
which is the systematic error currently found in cosmological
  analyses~\cite{amanullah2010,conley2011}. Therefore $\sigma_{w}$
will approach this value asymptotically and not improve beyond that.
The uncertainty from a self-calibrating fit, however, contains the
calibration error and will keep decreasing with larger numbers of
observed SNe. Figure \ref{fig:w-unc} shows uncertainties from
unperturbed and self-calibrating fits and the extrapolation of
$\sigma_{w}$ for larger surveys. Based on this, we find that for
number of observed SNe $N_{\textrm{SNe}} \mysim 3500$, a
self-calibrating fit will reach the same precision as the unperturbed
fit with systematics. Beyond that, our method will be outperforming
the conventional method of treating the uncertainty.

\section{Discussion and conclusion}
\label{sec:disc-concl}

In this paper we have applied the method of a self-calibrating Hubble
diagram to existing data for the first time to constrain uncertainties
in the flux measurements of the fundamental flux standard. To
construct a self-calibrating Hubble diagram, the SN light-curve data
is refit using SALT2 with a perturbed reference spectrum in which the
flux was increased or decreased slightly in a certain redshift range.
The change of the light-curve fit results due to this perturbation can
then be linearized and included in the Hubble fit, using the strength
of the perturbation as a nuisance parameter. 

Our analysis has shown that a self-calibrating approach to
incorporating calibration uncertainties works well with the current SN
Ia data sets. The self-calibrating Hubble diagram can be constructed
for any compilation of data from various surveys as the photometric
measurements are compared relative to a single reference star.
However, we identified non-linear effects of the perturbation of the
reference spectrum as an issue for using this method with the current
data and included a quality cut for this in the data selection
framework accordingly. Also we have omitted the retraining of the SN
Ia light-curve template. Based on a related study this is only a 5\%
effect and hence will not alter the conclusions significantly.
However, for a future study of the self calibration method, it would
be appropriate to further investigate the role of the light curve
training.

Self-calibrating Hubble fits for the current data can already
constrain the uncertainties for the spectrum of BD+17$^{\circ}\!$4708
to almost the same level as found in the literature, where the
spectrum was calibrated using three white dwarfs. Only for shorter
wavelengths, i.e.\ the $g$-band, these constraints are not that tight
because the redshift range usable for self-calibration is limited.
When using more than one perturbation parameter in the fit,
degeneracies between the parameters can appear. Even though they
increase the penalty of self-calibration at the current number of SNe,
the uncertainties behave in a statistically sound fashion. Therefore
the uncertainties can still be expected to decrease when larger data
sets become available. Extrapolations from the results for current
data show that the self-calibrating Hubble diagram will outperform the
standard method when the number of observed SNe has grown by about a
factor of five, i.e.\ $\mysim 3500$ SNe are available. This number of
SN observation will be achieved by upcoming surveys such as
DES\footnote{http://www.darkenergysurvey.org/} and
LSST\footnote{http://www.lsst.org/lsst/}. In this prediction, several
simplifications were made which might influence the number of SNe that
will actually be required. Simply scaling the uncertainties,
implicitly assumes that the accuracy of the CMB and BAO priors will
also improve similarly within the next decade. This is justified as
surveys to improve CMB and BAO results are already underway, e.g.\
Planck\footnote{http://www.rssd.esa.int/planck/} or
BigBOSS\footnote{http://bigboss.lbl.gov/}. Also the quality of the
data taken by future surveys will be better than the average quality
of the data set used here. Using more SN light-curve data with smaller
uncertainties (especially at low or high redshifts) will constrain the
uncertainties of the primary reference star better than extrapolated
by simply scaling the uncertainties. Furthermore this work only used
SNe with spectroscopic redshifts but future surveys will also rely on
photometric redshift determinations. In that case, calibration
uncertainties will also affect the resulting redshifts. This effect
will have to be analysed with techniques similar to those presented
here.

When constraining the calibration uncertainties using the
self-calibration method, one makes use of the assumption that SNe Ia
are truly standardizable and therefore do not evolve with redshift.
More precisely, the method requires that the rest-frame colour
relations remain the same for all redshifts. However, a change in the
metallicity, for instance, can have a greater effect on the rest-frame
$U$-band than on redder bands (see e.g. \cite{snf2011}). This change
in the colour relations will be partially absorbed during the
self-calibration procedure, which potentially can lead to a wrong
interpretation of the resulting cosmological parameters and affect the
nuisance parameters in the self-calibrating fit. Consequently any
significant departure of a nuisance parameter from zero would need to
be investigated carefully for evolution effects, e.g. by separating
the SNe by host type and/or SN characteristics, as well as possible
survey specific effects that can be investigated by subdividing the
data according the surveys.

In our analysis we have only constrained the uncertainties of the SED
of BD+17$^{\circ}\!4708$. With the framework presented here, it is
also possible to constrain the zero points of individual surveys,
however, at the potential expense of more nuissance parameters. Once
sufficiently large SN data sets from single surveys become available
(e.g.\ $\sim10^5$ SNe from LSST), that do not require the combination
with other samples, perturbing the zero point will have the same
effect on the cosmological fit results as perturbing the reference
spectrum. Hence, it will not further worsen the constraints on $w$,
while absorbing another significant source of systematic error.

\section*{Acknowledgements}
\label{sec:acknowledgements}

We would like to thank A.\ Conley for helpful discussions.
The authors acknowledge support by the DFG through TR33 ``The Dark Universe''. 

\newpage
\bibliographystyle{JHEP}
\bibliography{./ref-short}

\end{document}